\documentclass[doublecol,aps,epl,preprintnumbers,nofootinbib,floatfix]{epl2}
\usepackage{graphicx,color,amsmath}

\newcommand {\vct}[1] {\mathbf {#1}}

\newcommand{\Av}[1]{{\bf #1}}

\newcommand {\bnabla} {\mbox{\boldmath$\nabla$}}

\def\ln{{\operatorname{ln}}}

\def\rmi{{\mathrm{i}}}

\def\bnabla{{\boldsymbol\nabla}}

\begin{document}

\title{Kirkwood-Shumaker interactions and general thermal fluctuation forces}
\author{Nata\v sa Ad\v zi\' c \inst{1} \and Rudolf Podgornik \inst{1}}
\shortauthor{N. Ad\v zi\' c \etal}
\shorttitle{Kirkwood-Shumaker interactions}

\institute{\inst{1} Department of Theoretical Physics, J. Stefan Institute, and
Department of Physics, Faculty of Mathematics and Physics, University of Ljubljana, 1000 Ljubljana, Slovenia.} 

\date{}

\abstract
{We present an extension of the Kirkwood-Shumaker (KS) theory of proton-fluctuation interactions to situations where the perturbation theory, usually invoked to derive these interactions, fails. In order to do that we formulate a generalized theory of fluctuation interactions with non-linear macroion (surface) free energy term that naturally leads to the long-range KS interactions, but can also be straightforwardly generalized to situations where the perturbation expansion ceases to be valid. It also allows for other surface free energy terms to be treated within the same formalism, enabling us to derive the complete fluctuation interaction with monopolar and dipolar fluctuations included, leading to the KS theory as a limiting form.}

\maketitle

\section{Introduction}

Charge regulation is an old concept introduced first in the '20s by Linderstr\'om-Lang and later invoked by Kirkwood and Shumaker \cite{KS1,KS2} within the context of a  statistical mechanical perturbation theory to derive effective attractive interactions between two charged proteins in ionic solution \cite{Lund}. Charge regulation refers to the case, where the effective charge on a macroion, e.g. protein  surface, responds to the local solution conditions, such as local $pH$, local electrostatic potential, salt concentration, dielectric constant variation and the presence of other charged groups. While in nanoscale interactions \cite{Roger} one often assumes constancy of surface macroion charge, in fact the charge state of the dissociable groups on the macroion surface always depends strongly on the acid-base equilibrium that defines the fraction of acidic (basic) groups that are dissociated and should be consistently included in any theoretical formulation \cite{Igal}. 

As was first recognized by Kirkwood and Shumaker theoretically \cite{KS1,KS2} and later verified experimentally \cite{Lund}, the acid-base equilibrium induces long range thermal fluctuation forces that decay much slower than the ubiquitous van der Waals interactions \cite{Pit}. These Kirkwood-Shumaker (KS) interactions depend on the capacitance quantifying the molecular charge fluctuations that can be obtained from the titration curve \cite{Lund}. While the concept of KS interactions is invoked standardly in the investigation of protein-protein interactions \cite{BoJ}, the connection between these fluctuation forces and other types of thermal fluctuation forces in Coulomb systems has not been explored systematically. Here we will establish an exact connection between the monopolar van der Waals-type fluctuation interactions and the KS interactions, formulating the latter in such a way that they are amenable to further generalizations for systems where the original derivation and its analogues fail to be applicable.

\section{Kirkwood-Shumaker interactions}
The original derivation \cite{KS2} can be recast into a somewhat different form that actually accentuates the role played by the capacitance of the two apposed charge distributions \cite{BoJ}. This can be seen as follows:  take a system of two point-like macroions with a Hamiltonian of the form $${\cal H}(e_{1}, e_{2}; \Av r, \Av r')= \frac12 C^{-1}_1 e_1^2 + \frac12 C^{-1}_2 e_2^2  + e_1 e_2 G(|\Av r_1 - \Av r_2|)$$where ${\cal G}(|\Av r_1 -\Av r_2|) = {1}/{4\pi \varepsilon\varepsilon_0|\Av r_1 - \Av r_2|}$ is the standard Coulomb  Green's function where $\epsilon_w$ is the dielectric permittivity of the aqueous solvent. Finally, $C_{1,2}$ are the capacitances of the two macroions, i.e. inverse second derivatives of the self energies ${\cal H}_0(1,2)$ of the two charges, dependent on the assumed geometry and dimension of the charge distribution,  with respect to the macroion charge 
$
{C^{-1}}_{1,2} = \frac{\partial^2 {\cal H}_0(1,2)}{\partial e_{1,2}^2}.
$
The free energy for this system due to (monopolar) charge fluctuations is obtained by integrating the Boltzmann weight of the Hamiltonian w.r.t unconstrained fluctuations in $e_1$ and $e_2$ and is thus obtained as
\begin{equation}
{\cal F}(R) =  - k_BT \log{\int\!\!\!\int de_{1} de_{2} e^{- \beta {\cal H}}} \simeq  - C_1C_2~ {\cal G}^2(R). 
\label{gfcjkzfs}
\end{equation}
where $ R = |\Av r_1 - \Av r_2|$. Obviously this interaction energy scales as $R^{-2}$ and in fact coincides exactly with the KS interactions \cite{KS2} if the definition of the capacitance is taken into account.

In what follows it will turn out to be useful to derive also the KS force per unit area between two planar surfaces at separation D, bearing fluctuating charge groups. In this case a Hamaker-type summation gives the interaction pressure as
\begin{equation}
p = \frac{F(R)}{S} = - \frac{\partial}{\partial D} \int_D^{\infty} 2\pi R ~dR ~{ {\cal F}(R)} \simeq D^{-1}.
￼￼\end{equation}
It is obvious that the decay is slower then in the case of van der Waals interactions \cite{Pit} stemming from dipolar fluctuations between either two semi-infinite media or two thin layers, that scale as $D^{-2}$ and $D^{-5}$
%\begin{equation}
%p = \frac{F(R)}{S} = - \frac{A(D)}{12 \pi D^2} \qquad {\rm and/or} \qquad 
%p = \frac{F(R)}{S} = - \frac{2 A(D) a^2}{\pi D^5},
%\end{equation}
respectively. The KS fluctuation forces correspond to monopolar fluctuations and thus follow a different scaling either between point particles, $|\Av r_1 - \Av r_2|^{-3}$, or between fluctuating surface layers, $ D^{-1}$, then in the case of dipolar fluctuations.

\section{Charge regulation and capacitance}

We now concentrate on two charged (planar) surfaces with an intervening Coulomb fluid that can be either counterion-only or a uni-univalent electrolyte. For this case the mean-field free energy can be written as \cite{Perspective}
\begin{equation}
{\cal F}[\psi_{MF}(\Av r)]= \int_V f(\psi_{MF}(\Av r))d^3\Av r  + \oint\sigma_0 \psi_{MF} ~d^2\Av r
\label{action}
\end{equation}
where $\psi_{MF} $ with no argument is the surface value of the mean potential, with the volume part of the free energy being
\begin{equation}
f(\psi_{MF}(\Av r)) = - \frac{\epsilon_w\epsilon_0}{2 } ~
\bnabla \psi_{MF}(\Av r)^{2} - p(\psi_{MF}(\Av r))
%k_BT \sum_i{\lambda'}_i\!\!\int \!\! e^{- \beta e_i \psi_{MF}(\Av r)}d^3\Av r 
\end{equation}
with $\psi_{MF}(\Av r)$ the mean-field electrostatic potential, and $p(\psi_{MF}(\Av r))$ the van't Hoff osmotic pressure of the mobile ion species defined as $p(\psi_{MF}(\Av r)) = k_BT \sum_i{\lambda}_i \exp{- \beta e_i \psi_{MF}(\Av r)}$, where  index $i$ describes the various mobile ionic species, e.g. counterions $i = 1$, salt $i =2$ etc. and ${\lambda}_i$ are their respective fugacities. Above we also assumed that the surface charge density residing on the bounding surfaces is fixed and equal to $\sigma_0$. Minimizing the above free energy with respect to $\psi_{MF}(\Av r)$ we then obtain the standard mean-field Poisson-Boltzmann equation, with a boundary condition $-{\bf n}\cdot \bnabla \psi_{MF} = \sigma_0/\epsilon_w \epsilon_0$. This latter equality being true only for a fixed surface charge density.

Assume now a more general form of the surface free energy \cite{Rudi1,Andelman} by making the substitution  $\oint_S \sigma_0 \psi_{MF} d^2\Av r \longrightarrow  \oint_S f(\psi_{MF} )d^2\Av r $,  where $f(\psi_{MF} )$ is in general a non-linear function of the local potential. This form can be derived by including a surface specific, short range interaction potential into the partition function that then decouples into a volume and a surface term, both of them non-linear in the local electrostatic potential. While the volume term has a universal van't Hoff form, the surface term depends on the model of the surface-ion interaction. 

Various models are conceivable.  Let us first consider a surface Coulomb lattice gas partition function \cite{lattice} 
\begin{eqnarray}
\kern-15pt & & f(\psi_{MF}) = \oint_S\sigma_0 \psi_{MF} d^2 \Av r + \nonumber\\
& & +k_BT~ n_S \oint_S  \ln{\left( 1 + e^{- \beta \mu_S - \beta e_0 \psi_{MF}} \right)} d^2 \Av r,
\label{actions}
\end{eqnarray}
where $\sigma_0 = e_o n_S$. In the argument of the $log$ we can recognize the partition function for a system that corresponds to a de-protonated state as the ground state, and a protonated state with an effective energy $\beta \mu_S = \ln{10}(pH - pK)$ corresponding to the chemical energy of protonation. $n_S$  is the density of the dissociation sites on the surface.  

In fact the above free energy corresponds exactly to the Ninham-Parsegian site-dissociation model \cite{Pars-CR,Igal}, described with the de-protonation equilibrium $\rm AH \leftrightarrow A^{-} + H^{+}$, with an equilibrium constant $ \rm K = {[H^+]_S [A^-]}/{[AH]}$.
%\frac{\alpha}{1 - \alpha} = [H^+]  e^{- \beta e_0 i\psi_{MF}(\Av r)} \frac{\alpha}{1 - \alpha} \end{equation}
The connection between pH at the surface, $-\log{[H^+]_S}$, and in the bulk, $-\log{[H^+]}$, is then given by $ -\log{[H^+]_S} = -\log{[H^+]} + \beta e_0 \psi_{MF}$, with $\rm pK = - \log{K}$ and $ \rm pH = - \log{[H^+]} $ \cite{gfhjew}. From here we obtain the charge regulation boundary condition \cite{Pars-CR}
\begin{equation}
- \epsilon\epsilon_0 \frac{\partial \psi_{MF}}{\partial \Av n} = \sigma(\psi_{MF})
= \frac{\partial f(\psi_{MF})}{\partial \psi_{MF}} 
\label{cčsrn}
\end{equation}
or explicitly
\begin{equation}
\sigma(\psi_{MF})  = \frac{\sigma_0}{2} \left( 1 + \tanh{{\textstyle\frac12} \left( \beta \mu_S + \beta e_0 \psi_{MF}\right)}\right),
\end{equation}
where the macroion charge $\sigma(\psi_{MF}) $ is regulated by the surface potential, itself depending on the parameters characterizing the solution, that allows the surface charge to span the interval $\sigma(\psi_{MF}(\Av r)) \in [0, \sigma_0]$

Other surface free energies are also possible \cite{Andelman,Dean} that can capture the charge regulation of proteins better then the above acid-base equilibrium model. In fact an obvious generalization of Eq. \ref{actions} would be
\begin{eqnarray}
\kern-5pt & & f(\psi_{MF}(\Av r)) = \oint_S\sigma_0 \psi_{MF} d^2 \Av r +  \nonumber\\
& & +~k_BT~n_S \oint_S  \ln{\left( 1 + e^{\beta \mu_S - \beta e_0 \psi_{MF}} \right)} d^2 \Av r.
\label{actions2}
\end{eqnarray}
where now $\sigma_0 \neq e_o n_S$. Again in the second term we have the partition function for a system with a protonated ground state, and a de-protonated dissociated state with an energy $\beta \mu_S + \beta e_0 \psi_{MF}$. From Eq. \ref{cčsrn} the surface charge density then follows  as
\begin{eqnarray}
%- \epsilon\epsilon_0 \frac{\partial \psi_{MF}(\Av r)}{\partial \Av n} =  
\kern-5pt & & \sigma(\psi_{MF}) = \left( \sigma_0 - {\textstyle\frac12} e_0 n_S\right) + \nonumber\\
& & + {\textstyle\frac12} e_0 n_S \tanh{{\textstyle\frac12} \left( -\beta \mu_S + \beta e_0 \psi_{MF}\right)}
\end{eqnarray}
and the macroion charge is regulated in the interval $\sigma(\psi_{MF}) \in [\sigma_0 - e_0 n_S, \sigma_0]$. This would correspond to a coarse-grained description of a protein with an average value of the pK and a pH dependent positive or negative net charge. One should of course realize that not only protons, but any binding of solution ions defining the acid-base equilibrium of the surface charges could be described within the same model with appropriate changes in the definition of the binding energy.

From the general model Eq. \ref{actions2} we can ￼￼now calculate the corresponding surface capacitance that is defined as
\begin{equation}
{\cal C}_S = \frac{\partial^2 f(\psi_{MF})}{\partial (\beta e \psi_{MF})^2}  %= k_BT~n_S ~ \frac{ e^{ \ln{10}(pH - pK) + \beta e_0 \psi_{MF}(\Av r)} }{\left(1 + e^{ \ln{10}(pH - pK) + \beta e_0\psi_{MF}(\Av r)}\right)^2 } 
= {\textstyle\frac12}  (\beta e_0^2) n_S \left( 1 - \left( \frac{\sigma(\psi_{MF})  - \tilde\sigma_0}{{\textstyle\frac12} e_0 n_S}\right)^2\right).
\label{bdjs}
\end{equation}
with $\tilde\sigma_0 = \sigma_0 - {\textstyle\frac12} e_0 n_S$.
%Alternatively one can also write ${\cal C}_S = \frac{\partial \sigma}{\partial (\beta e \psi_{MF}(\Av r))} = \frac{\partial \sigma}{\partial (\ln{10}~pH)}$ since the dependence $\sigma(\psi_{MF}(\Av r))$ is the same as $\sigma( \log{10} pH)$. 
At the point-of-zero-charge (PZC), where $\sigma(\psi_{MF}) = 0$ the capacitance has a maximum at $\sigma_0 = {\textstyle\frac12} e_0 n_S$. By regulating the pH for each separation between surfaces one can in fact find the corresponding PZC at that separation, making the fluctuation contribution to the free energy dominant. 

\section{Fluctuation interactions}

We now evaluate the Gaussian fluctuations around the above mean-field solution. This can be done rather straightforwardly by exploiting the exact  field theoretical represenation of the Coulomb fluid partition function \cite{funint,Netz-Orland}, identifying the mean-field PB solution as its saddle-point and then expanding the local fluctuating potential up to the second order \cite{Rudi1}. 

The field-theoretic representation of the partition function of a Coulomb fluid with pure electrostatic interactions can be obtained from an integral over all fluctuating potential profiles of the field action ${\cal S}[\phi(\Av r)] =  {\cal F}[\psi_{MF}(\Av r)\longrightarrow i \phi(\Av r)]$, where $\phi(\Av r)$ is the fluctuating potential and $i$ is the imaginary unit (for details see Ref. \cite{Perspective}).  Here ${\cal F}$ refers to the complete, volume plus surface part, free energy. The total free energy is then obtained in the form of a functional integral over the fluctuating potential
\begin{equation}
{\cal F}(D) = \int {\cal D}[\phi(\Av r)] e^{- \beta {\cal S}[\phi(\Av r)] }
\end{equation}
and can be decomposed as ${\cal F}(D)= {\cal F}[\psi_{MF}]  + {\cal F}_2(D)$, where 
\begin{equation}
{\cal F}_2(D) = {\textstyle\frac{1}{2}} {\rm Tr} \log{ \frac{\delta^2 {\cal F}[i \phi(\Av r)]]}{\delta \phi({\vct r}) \delta \phi({\vct r}')}\bigg\vert_{\phi=-\rmi \psi_{MF}} }
\end{equation}
by considering only Gaussian deviations (one-loop correction) from the MF free energy, 
%The partition function for Gaussian fluctuations around the mean-field is then  
%\begin{equation}
%\Xi = \int {\cal D}[\phi(\Av r)] \exp{\left( -\beta {\cal F}[\psi_{MF}] + {\textstyle\frac{1}{2}} \int {\mathrm{d}}{\mathbf r}\,{\mathrm{d}}{\mathbf r}'\,  \varphi({\vct r}) \varphi({\vct r}')\, \frac{\delta^2 \beta {\cal F}[i \phi(\Av r)]]}{\delta \phi({\vct r}) \delta \phi({\vct r}')}\bigg\vert_{\phi=-\rmi \psi_{MF}} \!\!\!\!+ {\cal O}(\varphi^3)\right)}
%\end{equation}
%These fluctuations then lead to Gaussian (one-loop) correction to the MF free energy  ${\cal F}(D)= {\cal F}[\psi_{MF}]  + {\cal F}_2(D), $ 
analogous to vdW-type interactions  \cite{funint,Netz01,Netz-Orland}. Assuming a constant MF potential the fluctuation contribution to the free energy, ${\cal F}_2(D)$, can be obtained explicitly. Limiting ourselves to the case of two planar surfaces of area $S$ and dielectric response function $\varepsilon'$ at separation $D$ we end up with
\begin{equation}
\frac{{\cal F}_2(D)}{S} = \frac{k_BT}{4 \pi} \int_0^{\infty} Q dQ \log{\left( 1 - \Delta_{12}^2(Q) e^{-2 D p}\right)},
\label{eq:klfghjdf}
\end{equation}
%\begin{eqnarray}
%{\cal F}_2 &=& \frac12 k_BT \sum_Q~ \log{\left( 1 - \Delta_{12}^2(Q) ~e^{- 2 p D }\right)},
%\label{bgcfhjsik}
%\end{eqnarray}
where the sum is over 2D longitudinal wave vectors and  
\begin{equation}\label{eq:beast}
 \Delta_{12}^2(Q)  =  \frac{\left( {\cal C}_{S_1} + \beta (\epsilon' Q - \epsilon p)\epsilon_0\right)}{\left( {\cal C}_{S_1} + \beta (\epsilon' Q + \epsilon p)\epsilon_0\right)} \frac{\left( {\cal C}_{S_2} + \beta (\epsilon' Q - \epsilon p)\epsilon_0\right)}{\left( {\cal C}_{S_2} + \beta (\epsilon' Q + \epsilon p)\epsilon_0\right)} .
\end{equation}
Above, the surface capacitances ${\cal C}_{S_1}, {\cal C}_{S_2}$ refer to the two bounding surfaces and $D$ is the separation between the surfaces. We assumed that the region between the surfaces is filled with a uni-univalent salt with inverse Debye length $\kappa$, with $p^2 = Q^2 + \kappa^2$.  While a detailed analysis of the case where the intervening mean potential is not zero can be done \cite{Rudi1}, we do not explore this venue since we want to concentrate on conceptual aspects relegating the computational details for later endeavours.

Without any surface capacitance, ${\cal C}_{S_1}, {\cal C}_{S_2} = 0$, i.e.assuming that the surface free energy is linear in the surface potential,  
%\begin{eqnarray}
 %\Delta_{12}^2(Q)  &=&  \Big(\frac{\epsilon' Q - \epsilon p}{\epsilon' Q + \epsilon p} \Big)^2,\label{eq:flim}
%\end{eqnarray}
the fluctuation contribution Eq. \ref{eq:klfghjdf} is reduced to the zero frequency Lifshitz term in the full expression of the vdW interaction, as it should \cite{Pit}.  Another interesting limiting case is no dielectric discontinuity $\epsilon'  = \epsilon$ and salt on both sides of the surface, which would correspond to the original KS model that does not consider any dielectric discontinuities.
%Here we derive
%\begin{eqnarray}
 %\Delta_{12}^2(Q)  &=& \frac{ {\cal C}_{S_1} }{\left( {\cal C}_{S_1} + 2 \beta \epsilon \epsilon_0 ~p\right)} \times \frac{ {\cal C}_{S_2}}{\left( {\cal C}_{S_2} + 2 \beta \epsilon \epsilon_0~p\right)} = \Big(\frac{ {\cal C} }{{\cal C} + 2 \beta \epsilon \epsilon_0 ~p} \Big)^2,\label{eq:seclim}
%\end{eqnarray}
Assuming that the two surfaces have identical properties, i.e. ${\cal C}_{S_1} = {\cal C}_{S_2} = {\cal C}_S = (\beta e_0^2)~ n_S~ c$, the fluctuation interaction between two point-like macroions with dissociable sites, as analyzed by KS, can then be obtained from the planar case, Eq. \ref{eq:klfghjdf}, via the inverse Hamaker summation that leads to
\begin{equation}
\lim_{n_{S1}, n_{S2} \longrightarrow 0}\frac{\partial  }{\partial D} \left( \frac{{\cal F}_2(D)}{S}\right) = 2\pi ~n_{S1} n_{S2} ~D g_2(D),
\end{equation}
where $g_2(D)$ is now the fluctuation interaction free energy between two point-like macroions at separation $D$. In the limit of small $n_S$  we then derive
%\begin{equation}
%\lim_{n_{S1}, n_{S2} \longrightarrow 0}\frac{\partial^2  }{\partial D^2} \left( \frac{{\cal F}_2(D)}{S}\right) \simeq 
%- \frac{k_BT}{4 \pi} \int_0^{\infty} Q dQ  \Delta_{12}^2(Q) e^{-2 D \sqrt{Q^2 + \kappa^2}} 
%= - \frac{k_BT}{4 \pi} \frac{4~ {\cal C}_{S_1} {\cal C}_{S_2}}{(2 \beta \epsilon\epsilon_0)^2} \int_0^{\infty} Q dQ   ~e^{-2 D \sqrt{Q^2 + \kappa^2}} .
%\end{equation}
the KS interaction in the form
\begin{equation}
g_2(D) = - \frac{\pi {k_BT}}{8} ~\ell_B^2~{  { c}^2} \frac{e^{- 2\kappa D}}{D^2},
\end{equation}
where $\ell_B$ is the Bjerrum length. $g_2(D) $ indeed scales as $D^{-2}$ for vanishing screening. The KS interactions obtained in this limit are thus just monopolar thermal vdW fluctuation interactions% apart from the fact that in general when evaluating the capacitances they need to be taken at the proper surface potential value
. This reformulation is necessary if one wants to generalize the concept of KS interactions to strongly coupled systems or indeed to systems that would be described with different forms of the non-linear surface free energy.

In general it is clear from Eq. \ref{eq:klfghjdf}  that there is a competition between monopolar and dipolar fluctuations. 
%In  fact in the case of no intervening electrolyte, that is no screening, Eq. \ref{eq:klfghjdf} can be written as 
%\begin{equation}
%\frac{{\cal F}_2(D)}{S} = \frac{k_BT}{4 \pi (2D)^2} \int_0^{\infty} u ~du \log{\left( 1 - \left( \frac{ 2D ~{\cal C}_{S} + \beta \epsilon_0 (\epsilon - \epsilon') u }{2D ~{\cal C}_{S} +  \beta \epsilon_0 (\epsilon + \epsilon') u} \right)^2 e^{-u}\right)},
%\label{eq:klfghjdf1}
%\end{equation}
Ignoring the screening effects different terms in the fluctuation interaction pressure, $p = - \frac{\partial }{\partial D} (\frac{{\cal F}_2(D)}{S})$, respectively scale as ${\cal C}_S^2 /D$, ${\cal C}_{S} \Delta /D^2$ and $\Delta^2 /D^3$, corresponding to monopolar,  coupled monopolar-dipolar and standard dipolar fluctuations (zero frequency) vdW  terms \cite{Pit}. While the monopolar and the dipolar terms are always attractive, the mixed term depends on the sign of the dielectric missmatch $\Delta = (\epsilon - \epsilon')/ (\epsilon + \epsilon')$. At large separation it is always the monopolar term that dominates.

As we already indicated we did not specifically take into account the fact that there is usually a mean potential profile between the interacting surfaces and one would in fact need to evaluate the capacitances Eq. \ref{bdjs} at the surface value of this potential $\psi_{MF}(\Av r)$, i.e., ${\cal C}_S = \frac{\partial^2 f_S(\psi_{MF})}{\partial (\beta e \psi_{MF})^2} = {\cal C}_S (\psi_{MF})$. While this is one of the aspects that is usually not taken into account in the KS-like theories of charge fluctuation interactions between proteins \cite{Lund}, it has been taken into account numerically in analysis of the acid-base macroion charge equilibria \cite{Igal} but not of the corresponding charge fluctuation interactions.

\section{Generalizations}

It is not straightforward to incorporate dielectric discontinuities into the effective KS interactions. The Hamaker-type summation can not be used directly because an effective dielectric medium permitting the application of the Pitaevskii {\sl Ansatz} \cite{Pit} can not be constructed. Nevertheless the functional integral representation with specific dissociation equilibrium can pave the way towards this goal.  

Imagine two point-like macroions at positions $\Av r = \Av R_{1,2}$, with an intervening uni-valent electroyte. The field-theoretical partition function in this case has the field action of the form
\begin{equation}
{\cal S}[\phi(\Av r)] = \beta V(i\phi(\Av R_{1})) + {\cal F}[i \phi(\Av r)] + \beta V(i\phi(\Av R_{2})),
\label{ncnfwlhwk}
\end{equation}
where $V(i\phi(\Av R))$ represents the coupling of the point-like macroion with the fluctuating field and thus plays a role analogous to the surface part of the free energy in the case of a planar macroion analyzed above. The volume part ${\cal F}[i \phi(\Av r)]$ has been introduced before, Eq. \ref{action}. Again the partition function of this system can be obtained from an integral over all fluctuating potential profiles $[\phi(\Av r)]$ of the field action. The specific form of the coupling between the particles and the fluctuating field  is assumed in the form 
\begin{eqnarray}
&&V(i\phi(\Av R)) = i ~e \phi (\Av R) - \frac{\alpha}{2} \bnabla \phi (\Av R)^2 + \nonumber\\
& & +~k_BT~ N_0 ~  \ln{\left( 1 + e^{-\beta \mu_S - i \beta e_0 \phi(\Av R)} \right)},
\label{bckrbyf}
\end{eqnarray}
where $\alpha$ is the excess polarizability of the macroions \cite{Demery}, $e$ is the net charge and $N_0$ is the number of dissociating sites. For a spherical particle of radius $a$ and effective permittivity $\epsilon$  in aqueous solvent with dielectric permittivity $\epsilon_w$, the standard result is $\alpha = 4\pi \epsilon_0 a^3 (\epsilon - \epsilon_w)/(\epsilon + 2 \epsilon_w)$. The above coupling between the field and the macroions takes into account the polarizability but, being valid for a point-like macroion, not the dielectric images. 

In order to derive the KS interactions in this case we proceed as follows. Making a small potential expansion of Eq. \ref{bckrbyf} we can write to the order ${\cal O}[\phi^3 (\Av r)]$
\begin{equation}
V[i\phi(\Av r)] \simeq i \int_V \rho(\Av r) \phi (\Av r) d{\Av r} - {\textstyle\frac12}  \int\!\!\!\!\int_V\!\!d{\Av r} d{\Av r'} \phi (\Av r)  \tilde C (\Av r, \Av r') \phi (\Av r') 
\label{fiual+}
\end{equation}
where we introduced
\begin{equation}
\rho(\Av r) = \sum_{i=1,2}\left( e - \tilde e\right) \delta(\Av r - \Av R_i)
\end{equation}
and $\tilde e = e_0 N_0 \frac{e^{-\beta \mu_S}}{\left( 1 + e^{-\beta \mu_S}\right) } $. With
\begin{equation}
\tilde C(\Av r, \Av r') = \sum_{i=1,2}\left( c  +  \bnabla \alpha \bnabla' \right) \delta(\Av r - \Av R_i) \delta(\Av r' - \Av R_i),
\label{ncfhkrse}
\end{equation}
we have 
\begin{equation}
c = (\beta e_0^2) N_0 e^{-\beta \mu_S}/(1 + e^{-\beta \mu_S} )^2. 
\label{aefnk}
\end{equation}
The field-action Eq. \ref{ncnfwlhwk} can now be expanded to the second order yielding finally 
\begin{equation}
{\cal S}[\phi(\Av r)] \simeq {\textstyle\frac12} \int\!\!\!\int_V  d\Av rd\Av r' \phi(\Av r) {\cal G}^{-1}(\Av r, \Av r') \phi(\Av r'),
\end{equation}
where 
\begin{equation}
{\cal G}^{-1}(\Av r, \Av r') =  {\cal G}_0^{-1}(\Av r, \Av r')+ \tilde C(\Av r, \Av r').
\end{equation}
Assuming again an intervening DH electrolyte we finally derive
\begin{equation}
{\cal G}_0^{-1}(\Av r, \Av r') = -\epsilon\epsilon_0 (\bnabla^2 - \kappa^2)\delta(\Av r - \Av r').
\label{fbceq}
\end{equation}
The field integral of $\exp{-{\cal S}[\phi(\Av r)] }$ is of a general Gaussian form and can now be evaluated by a variety of methods
yielding for the fluctuation part of the free energy ${\cal F}_2 =  - k_BT~{\textstyle\frac12} {\rm Tr}\log{{\cal G}^{-1}(\Av r, \Av r')}$.
Retaining only the part that depends on the separation between particles 1 and 2, assumed to have identical properties, while taking into account the symmetry of the Green's function w.r.t. its arguments, we end up with 
\begin{equation}
\beta {\cal F}_2(\Av R_1, \Av R_2) =  - {\textstyle\frac12}\!\!\left( c ~{\cal G}_0(\Av R_1, \Av R_2) + \alpha \bnabla\bnabla' {\cal G}_0(\Av R_1, \Av R_2)\right)^2 + \dots,
\label{fhcklrwk}
\end{equation}
an obvious generalization of Eq. \ref{gfcjkzfs}, where ${\cal G}_0(\Av R_1, \Av R_2)$ is the DH Green's function corresponding to the kernel Eq. \ref{fbceq}. The two limits $\alpha \longrightarrow 0$ and $c \longrightarrow 0$ give the KS interaction and the zero frequency vdW interactions \cite{Ninham} respectively, as they should. To the above free energy one should also add the repulsive contribution of the charge-charge interaction stemming from the first term in the expansion Eq. \ref{fiual+}, to the lowest order identical to DH interactions between two charges of magnitude $\tilde e$ located at $\Av R_1, \Av R_2$ and is as such not particularly interesting.

A more important generalization concerns the fluctuating dipole contribution. One could in fact introduce KS interaction of type 1 (monopole proton charge fluctuations) \cite{KS2} and KS interaction of type 2 (dipole moment fluctuations) \cite{KS1}, both in fact due to proton fluctuations at the dissociation sites of the macroion. The dipole moment fluctuation contributes a remarkably large increment to the dielectric constant of water, that is usually in fact negative (decrement) for simple salts \cite{decrement}.

Dipolar fluctuations would be described by a modified Boltzmann weight in  Eq. \ref{bckrbyf}
\begin{eqnarray}
e^{- i \beta e_0 \phi(\Av R)}& &  \longrightarrow \mathopen< e^{- i \beta e_0 \phi(\Av R) - i {\bf p}\cdot \bnabla \phi(\Av R)} \mathclose>_{\Omega} = \nonumber\\
\kern-10pt & & e^{- i \beta e_0 \phi(\Av R)} \frac{\sin{\beta p \vert \bnabla \phi(\Av R)\vert}}{\beta p \vert \bnabla \phi(\Av R)\vert}.
\end{eqnarray}
The above average was performed with respect to the internal (orientational) coordinates ${\Omega}$ describing the polarization fluctuations of the macroion with $p \sim e_0 a$, where $a$ is the radius of the macroion. Expanding to the second order in the potential gradient we obtain Eq. \ref{ncfhkrse} but with a redefined polarizability
\begin{equation}
\alpha \longrightarrow 4\pi \epsilon_0 a^3 \frac{(\epsilon - \epsilon_w)}{(\epsilon + 2 \epsilon_w)} + {\textstyle\frac13} (\beta p)^2 \frac{e^{-\beta \mu_S} }{1 + e^{-\beta \mu_S} }.
\label{cwrtyqes}
\end{equation}
The fluctuating dipole moment thus affects the standard vdW interactions stemming from the polarizability of the macroion that now contains the sum of excess polarizability of the macroion (negative) and the dipole moment fluctuations (positive). For proteins, with a large dielectric increment, the latter is much larger than the former and sets the overall sign of the total polarizability.

The total fluctuation interaction, Eq. \ref{fhcklrwk}, is then composed of the fluctuating monopolar charge contribution, codified by the capacitance Eq. \ref{aefnk}, and the fluctuating dipole contribution, described by the total excess polarizability Eq. \ref{cwrtyqes} and is as a whole dominant when the average charge is vanishing, i.e. close to the PZC of the macroion. Obviously this {\sl generalized KS interaction} contains the monopolar and dipolar fluctuation contributions in a highly non-additive way and could in fact go through a local maximum as a function of the separation $\vert \Av R_1- \Av R_2\vert$. Pairwise additive forms \cite{Ninham} are highly approximate and in general not valid.

\section{Conclusions}

We introduced generalized KS interactions based on the field-theoretic formulation that allows for inclusion of a macroion surface free energy describing the charge dissociation processes. This surface free energy is a straightforward generalization of the charge regulation formalism \cite{Pars-CR}  introduced specifically within a mean-field Poisson-Boltzmann context. For small macroions this charge regulation surface free energy then tends towards a point-like non-linear form that gives a non-zero capacitance which sets the magnitude of the monopolar charge fluctuations. The specific advantage of the present formulation is that it shows clearly that KS interactions (of type 1) stem directly from the monopolar charge fluctuations, in distinction to the standard vdW interactions which are dipolar in nature. It is for this reason that they are of a fundamentally longer range. Furthermore, the monopolar and dipolar fluctuation components contribute non-additively to the total fluctuation interaction and could in principle depend non-monotonically on the separation between interacting macroions. 

The formulation of the generalized KS interactions based on specific forms of the charging equilibria of macroions and its conversion into the functional integral representation opens up new venues for a straightforward generalizations of the KS theory such as e.g. strong coupling limit \cite{Perspective} that were not envisioned in the original formulation or its progenies. Though, when referring to macroions, we were specifically motivated by protein interactions, the presented analysis remains valid for any type of charged colloids with acid-base surface equilibria.

\section{Acknowledgment}

R.P. would like to thank the hospitality of Prof. R.R. Netz during his stay at the Freie Universit\" at Berlin and the Technische Universit\" at Berlin as a visiting professor.


\begin{thebibliography}{99}

\bibitem{KS1} J. Kirkwood and J.B. Shumaker,  Proc. Natl. Acad. Sci. USA  {\bf 38} 855  (1952).
\bibitem{KS2} J. Kirkwood and J.B. Shumaker, Proc. Natl. Acad. Sci. USA  {\bf 38} 863 (1952). 
\bibitem{Lund} M. Lund and B. J\" onsson, Quart. Rev. Biophys. {\bf 46} 265-281 (2013).
\bibitem{BoJ} M. Lund and B. J\" onsson, Biochemistry {\bf 44} 5722 (2005). F.L. B. da Silva, M.  Lund, B. Jonsson, and T. \AA kesson, J. Phys. Chem. B  {\bf 110} 4459 (2006). F.L. B. da Silva and B. Jonsson, Soft Matter, {\bf 5} 2862 (2009). 

\bibitem{Roger} R. French et al., Rev. Mod. Phys. {\bf 82}, 1887 (2010).
\bibitem{Igal} M. J. Uline, Y. Rabin, and I. Szleifer, Langmuir {\bf 27}, 4679 (2011). G. S. Longo, M. Olvera de la Cruz and I. Szleifer, Macromolecules {\bf 44}, 147 (2011). 
%\bibitem{Tima} S. Timasheff, H. Dintzis, J. Kirkwood \& B. Coleman, Proc. Natl. Acad. Sci. USA {\bf 41} 710 (1955).
\bibitem{Pit} V. A. Parsegian, Van der Waals Forces, Cambridge University Press, Cambridge (2005).

\bibitem{Pars-CR} B.W. Ninham and V.A. Parsegian, J. Theor. Biol. {\bf 31} 405 (1973).


\bibitem{Perspective} A. Naji, M. Kandu\v c, J. Forsman, and R. Podgornik, J. Chem. Phys. {\bf 139} 150901 (2013). 

\bibitem{lattice} I. Borukhov, D. Andelman, H. Orland, Electrochimica Acta {\bf 46} 221 (2000).
\bibitem{funint} R. Podgornik, B. Zeks, J. Chem. Soc., Faraday Trans. II {\bf 84} 611 (1988). 
\bibitem{Netz-Orland} R.R. Netz and H. Orland, ￼Eur. Phys. J. E {\bf 1} 203 (2000)
\bibitem{Rudi1} R. Podgornik, J. Chem. Phys. {\bf 91} 5840 (1989).
%\bibitem{Podgornik89b}  R. Podgornik, J. Phys. A: Math. Gen. {\bf 23} 275 (1990). 
\bibitem{Netz01} R.R. Netz, Eur. Phys. J. E {\bf 5} 557 (2001).
\bibitem{gfhjew} or in general minus log of the proton activity.
\bibitem{Andelman} D. Ben-Yaakov, D. Andelman, R. Podgornik, D. Harries, Curr. Op. Coll. \& Interf. Sci. {\bf 16} 542 (2011).
\bibitem{decrement} D. Ben-Yaakov, D. Andelman, R. Podgornik, J. Chem. Phys. {\bf 134} 074705-1 (2011).
\bibitem{Dean} D. S. Dean, and R. R. Horgan, Phys. Rev.E, {\bf 65} 061603 (2002).
\bibitem{Demery} V. D\' emery, D. S. Dean, and R. Podgornik, J. Chem.Phys. {\bf 137} 174903 (2012).
\bibitem{Ninham}  J. Mahanty and B.W. Ninham, {\sl Dispersion Forces}, (Academic Press, London, 1976).

\end{thebibliography}
\end{document}